\documentclass{sig-alternate}

\usepackage{listings}
\usepackage{graphicx}
\usepackage{amsmath}
\usepackage{algorithm2e}
\usepackage{csquotes}

\begin{document}

\title{Redynis: Traffic-aware dynamic repartitioning\\for a distributed key-value store}

\numberofauthors{1} 
\author{
	\alignauthor Vineet John\\
	\affaddr{David R. Cheriton School of Computer Science}\\
	\affaddr{University of Waterloo}\\
	\affaddr{Waterloo, Ontario, Canada}\\
	\email{vineet.john@uwaterloo.ca}
}

\maketitle
\begin{abstract}
Most modern data stores tend to be distributed, to enable the scaling of the data across multiple instances of commodity hardware. Although this ensures a near unlimited potential for storage, the data itself is not always ideally partitioned, and the cost of a network round-trip may cause a degradation of end-user experience with respect to response latency. The problem being solved is bringing the data objects closer to the frequent sources of requests using a dynamic repartitioning algorithm. This is important if the objective is to mitigate the overhead of network latency, and especially so if the partitions are widely geo-distributed. The intention is to bring these features to an existing distributed key-value store product, Redis\cite{redis-website}.\\
\end{abstract}

\\

\section{Introduction}
The objective of this project (Redynis) is to design a shared-something distributed architecture atop an existing key-value store that dynamically repartitions tuples based on traffic metrics.\\

Redynis takes its roots in intelligent data placement, and in essence, attempts to delegate the data ownership to the distributed key-value store instance that is closest to the most frequent sources of a request, by implementing a web-service as an intermediary layer to the key-value store on each node.\\

\section{Problem Statement}
The aim is to solve the problem of having to make frequency remote requests for local node cache misses. This needs to be solved in a manner that allows for a more usage-heuristic based dynamic repartitioning of the tuples, and build a framework that intelligently repartitions tuples for a distributed key-value store.\\

The motivation of Redynis is three-fold:
\begin{itemize}
	\item Reduce the network latency by dynamic repartitioning of the key-value tuples based on usage-traffic heuristics i.e. maximize the number of hits on the local data store.
	\item Leveraging the same usage-traffic heuristics to selectively purge stale data.
	\item Optimizations need to be non-blocking so as to not interfere with the regular execution of fetch requests.
\end{itemize}

Redynis is built with the purpose of implementing these features to reduce cross-node request latency.

\section{Similar Work}
This section lists the previous work done to solve the same or a similar problem to the one described in the previous section.\\

Attempts to identify ideal methods to dynamically partition data already exist. Two of them, SWORD\cite{quamar2013sword} and AdaptCache\cite{asad2016adaptcache}, rely on hyper-graphs to model the database workloads, and base the repartitioning decisions off this model. SCHISM\cite{curino2010schism} relies on graph partitioning to find a predicate-based explanation for the ideal partitioning strategy. E-store\cite{taft2014store} relies of a strategy of skewed placement, where the data placement decision is taken based on usage heuristics, while avoiding replication. \\

Redynis, however, is implemented in a manner that avoids expensive graph traversals and is able to log usage heuristics and perform a usage analysis for a key in constant time. In addition, since Redis is widely used in a lot of industrial tech stacks\cite{redis-popularity}, it makes for an easier deployment strategy, compared to switching over to a new system. Also included, is a web API for ease of data access, which minimizes the development overhead of having to implement a language specific client to interact with the key-value store. Any other key-value store can be also swapped in, in place of Redis, without any changes to the client's view of the architecture.\\

\section{Model Assumptions}
The following assumptions are made, with respect to the system architecture and the problem statement:
\begin{itemize}
	\item The load balancing layer on the application servers hosting the web-service ensures that the request from clients is served by the application server closest to the client. This problem has already been solved by host resolution techniques by a DNS setup\cite{pan2003overview}.
	\item The nature of the workload, as is the case with most key-value stores, is predominantly read requests.
	\item The network of nodes is geo-distributed
	\item Minimal memory usage on each of the nodes is a desirable property
	\item $\displaystyle size(value) \gg size(key)$
\end{itemize}

\section{System Architecture}
\begin{figure*}
\centering
\includegraphics[width=\textwidth]{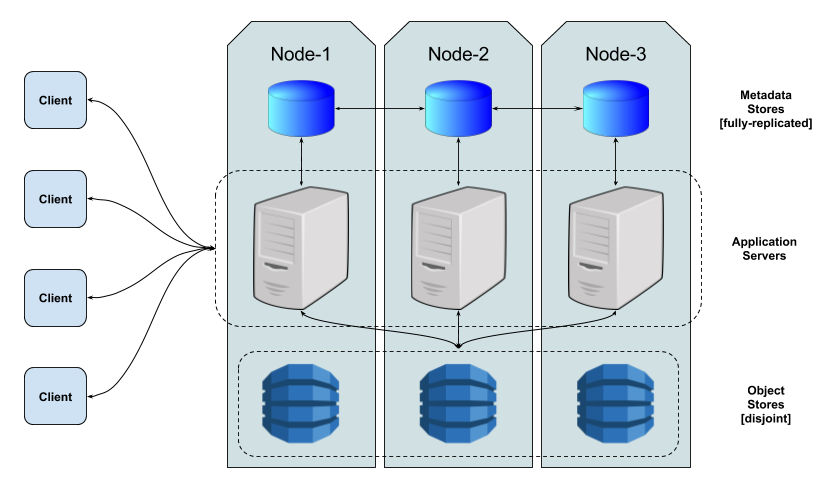}
\caption{System Architecture}
\label{fig:sysarch}
\end{figure*}

This section elaborates on the components the system is comprised of, and explains the points of interaction between them. \textbf{Figure \ref{fig:sysarch}} shows a graphical representation of the architecture.\\

\subsection{Components}
The components of the architecture are listed below:
\begin{itemize}
	\item \textbf{Web service layer:} This layer of abstraction over the key-value store, is deployed on the application server nodes. It receives requests, reads placement details from the metadata layer and acts accordingly. 
	\item \textbf{Data Layer:} This is the underlying in-memory datastore which the objects are primarily stored in. There is a single key-value store instance running on each of the nodes.
	\item \textbf{Metadata Layer:} This is a smaller key-value store for metadata, which is a separate cluster running on the same nodes as the actual key-value stores. It stores the key metadata, like current placement, usage heuristics and recency of access.
	\item \textbf{Placement Daemon:} This continuous process keeps running offline in periodic intervals to repartition the keys, based on the placement strategy described in \textbf{Algorithm \ref{algo-place}}.
\end{itemize}

\subsection{Component Interaction}
Component Interaction can be enumerated as below:
\begin{itemize}
	\item The web service deployment instances are agnostic of each other.
	\item The metadata layer and the caching layer are agnostic of each other.
	\item A given web-service on node can initiate a key-value store request call on any of the instances in the cluster of nodes.
	\item The data placement daemon is agnostic of the web-server layer. It merely reads from the metadata layer and enforces changes to the key-value store instances.
\end{itemize}

\section{Key concepts}
This section contains a description of the key concepts that the dynamic repartitioning strategy, is heavily dependant on, including the `Ownership coefficient' and the data format in which metadata for each key is maintained.\\

\subsection{Ownership coefficient} \label{Ownership-coefficient}
The ownership coefficient (H) determines which nodes need to have a local copy of a particular key/object.\\

During the analysis phase of the data placement daemon, the key usage for each node is calculated using equation \ref{eq:1}:
\begin{equation*}g(O, x) = count(accesses\ on\ object\ O)\ by\ node\ x \end{equation*}
\begin{equation} \label{eq:1} f(O,x) = \frac{g(O, x)}{g(O, \forall nodes)} \end{equation}

If equation \ref{eq:2} holds true, then node `x' is deemed eligible to possess a replicated copy of object O.
\begin{equation} \label{eq:2} f(O, x) - H \geq 0 \end{equation}

The above conditions operate under the constraint defined in equation \ref{eq:3}
\begin{equation} \label{eq:3} H - \frac{1}{n} \leq 0 \end{equation}
where `n' is the number of nodes in the architecture. This constraint is defined to avoid host starvation of key ownership, especially for cases in which hosts might have close to equivalent access metrics, and result in undesired deletion of keys from nodes.\\

\subsection{Metadata format}
The data object used to store the metadata for each tuple is given below.
\begin{lstlisting}
{
    `totalAccessCount': 17,
    `hosts': [
        `node-1',
        `node-3'
    ],
    `hostAccesses': {
        `node-1': 9,
        `node-2': 3,
        `node-3': 5
    },
    `lastAccessedDate': 1480725771235
}
\end{lstlisting}
\textit{hosts} is a hashed set, \textit{hostAccesses} is a data-dictionary, and the numeric values are positive integers. \textit{lastAccessedDate} denotes when the key in question was last accessed, in terms of milliseconds elapsed since the epoch.\\

\section{Algorithmic Approach}
This section describes the algorithms being used to implement the architecture, including fetching, storing and repartitioning tuples.\\

\subsection{Fetching tuples}
Described in \textbf{Algorithm \ref{algo-fetch}}.\\

\begin{algorithm}
\KwData{key}
\KwResult{value/null}
\DontPrintSemicolon 
\;
query metadata data for key\;

\uIf{$metadata == null$}{
	return null;
}
\uElse{
	$owner\_hosts = metadata.hosts$\;
	\uIf{$current\_host \in owner\_hosts$}{
		make local request to get data\;
	}
	\uElse{
		make remote request\\
		(incurring additional latency)\;
	}
	spawn async thread and collect access metrics\;
	return value to user\;
}
\;
\label{algo-fetch}
\caption{Fetching values}
\end{algorithm}

\begin{algorithm}
\KwData{key, value}
\KwResult{success = true/false}
\DontPrintSemicolon 
\;
query metadata data for key;\\

\uIf{metadata == null}{
	store new key and value locally\;
	generate metadata object\;
	post metadata object to metadata layer\;
}
\uElse{
	$owner\_hosts = metadata.hosts$\;
	\uIf{$current\_host \in owner\_hosts\ \land$\;$length(owner\_hosts) == 1$}{
		key is only present at current-host\;
		post value locally\;
	}
	\uElseIf{current-host == write-serializer}{
		post value to owner-hosts\;
	}
	\Else{
		relay store request to write-serializer node\;
	}
}
\;
\uIf{no-exception-thrown}{
	return true\;
}
\uElse{
	return false\;
}
\;
\label{algo-store}
\caption{Storing values}
\end{algorithm}

\subsection{Storing tuples}
Described in \textbf{Algorithm \ref{algo-store}}.\\

\subsection{Repartitioning tuples}
Described in \textbf{Algorithm \ref{algo-place}}.\\

\begin{algorithm}
\KwData{master-metadata-host}
\KwResult{null}
\DontPrintSemicolon 
\;
initialize $H = ownership.coefficient$\;
initialize $owner\_hosts$\;
initialize $delete\_hosts$\;
\;
\For{$key \in all\_keys$}{
	$keyaccesses = key.metadata.totalaccesses$\;
	$current\_hosts = key.metadata.hosts$\;
	$owner.accessmap = key.metadata.accessmap$\;

	\uIf{$now > (key.metadata.hosts - expirytime)$}{
		delete $key$ from $current\_hosts$\;
		delete $key$ from metadata\;
	}

	\;
	\For{$hostaccess\_pair \in owner.accessmap$}{
		$ \displaystyle f = \frac{hostaccess\_pair.accesses}{keyaccesses}$\;
		\uIf{$f \geq H$}{
			add $hostaccess\_pair.host$ to $owner\_hosts$\;
		}
		\uElse{
			add $hostaccess\_pair.host$ to $delete\_hosts$\;
		}
	}
	\;
	$new\_hosts = owner\_hosts - current\_hosts$\;
	$obsolete\_hosts = current\_hosts \cap delete\_hosts$\;
	\;
	add new hosts and delete obsolete hosts\
	from metadata\;
}

\label{algo-place}
\caption{Placement Algorithm}
\end{algorithm}

\section{Test Setup}
This section describes the details of the test setup, including the test bed and the nature of the experiments conducted.

\subsection{Testbed}

\subsubsection{Service Infrastructure} \label{Service-Infrastructure}
The testing for this experiment was done on a cluster of 3 nodes, with 12 CPU cores and 16 GB of main memory.
Each of these nodes contains a deployment setup as follows:
\begin{itemize}
	\item RedynisService \cite{redynis-svc}
	\item Redis instance (as the actual key-value store)
	\item Redis instance (as the metadata store)
\end{itemize}

Of these, one of these nodes is configured to be the master propagator, in order to serialize write transactions and ensure correctness of value data across the Redis instances.\\

\subsubsection{Placement Infrastructure}
A single node will be running a continuous execution for RedynisDaemon \cite{redynis-daemon}
The node's hardware specifications the same as the nodes running the service infrastucture (\ref{Service-Infrastructure}).\\

\subsection{Workloads}
YCSB workloads for RESTful web-services was used to benchmark this experimental setup.\\

The tests run were permutations of the below configurations:
\begin{itemize}
	\item  Workload Read requests (\%) ranging from 100(all reads) to 50(write-heavy)
	\item  Uniform key-value access distribution vs Skewed key-value access distribution
\end{itemize}

The uniform distribution workload accesses all the tuples an equal amount of times, whereas the skewed distribution workload accesses is a zipfian distribution that requests 10\% of the data items 90\% of the time.\\

All of the workloads have been run on a uniform set of 100,000 total requests. To simulate a widely geo-distributed network of nodes, the incurred latency for making a request to a remote node is simulated to be 100ms\cite{latency-stats}, whereas there is no incurred penalty for making a local request.\\

\begin{figure*}[ht]
\centering
\includegraphics[width=\textwidth]{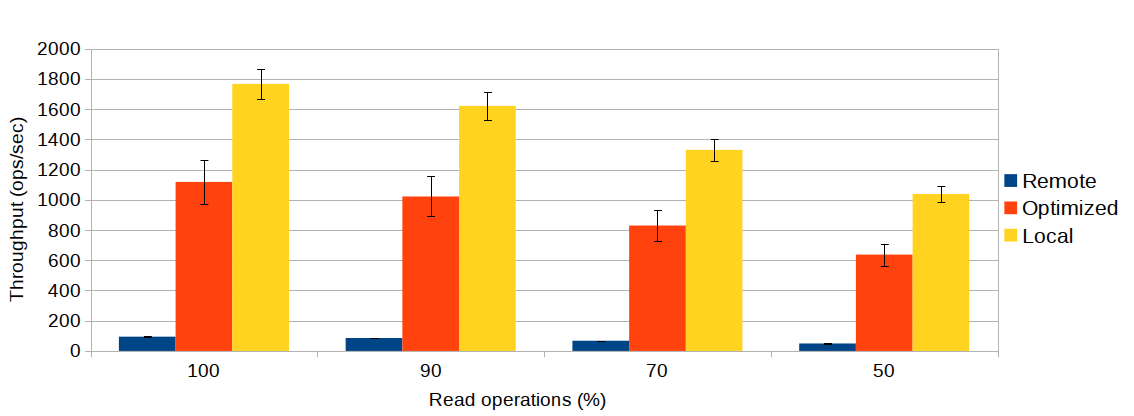}
\caption{Uniform Object Access Distribution}
\label{fig:res-unif}
\end{figure*}

\begin{figure*}[ht]
\centering
\includegraphics[width=\textwidth]{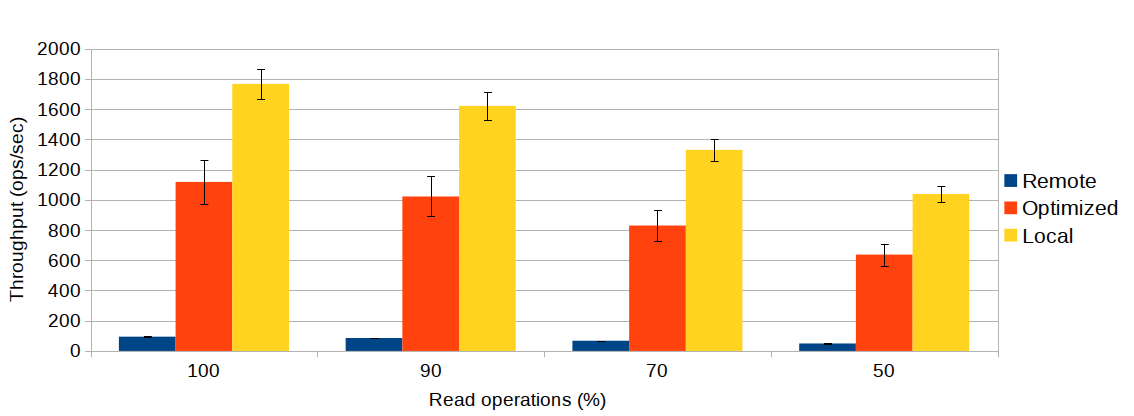}
\caption{Skewed Object Access Distribution}
\label{fig:res-skew}
\end{figure*}

\section{Experimental Results}

The experimental results for uniform distribution of object access and skewed distribution are indicated in \textbf{Figure \ref{fig:res-unif}} and \textbf{Figure \ref{fig:res-skew}} respectively.\\

Each of the bars plotted for throughput have additional error bars to indicate the 99\% confidence interval for the distribution across the multiple iterations of the experiment performed. \\

The different scenarios enumerated in the graphs are described below:
\begin{itemize}
	\item \textbf{Local:} All requests for keys made are served by the key-value store on the local node. This is the theoretically ideal scenario.
	\item \textbf{Remote:} All requests for keys made are served not available on the local key-value store, and for each request, the penalty of having to retrieve the key's value from a remote node, is incurred.
	\item \textbf{Optimized:} All requests for keys made are served not available on the local key-value store. However, as the requests keep being made, the usage statistics are logged, and the Redynis daemon, following the Ownership coefficient policy detailed in \textbf{section \ref{Ownership-coefficient}}, replicates the keys to the local key-value store on the fly, to mitigate the penalty incurred by having to make remote requests for a frequently accessed key.
\end{itemize}

The hypothesis being tested by the experiment is to examine if the optimized option is a good-enough alternative to a naive global replication of all keys across all nodes in the key-value cluster. The experimental results corroborate this hypothesis.\\

\section{Conclusions}
The performance of the system designed for the experiment is approximately ten-fold better than the scenario in which all the requests made are re-routed to remote nodes. It is also nearly comparable to the theoretical ideal key-value store that contains all the keys, which could prove to be a helpful alternative to having a fully-replicated Redis cluster. The experimental results point to the hypothesis being proven true. \\

The Redynis system expands the feature set of an already widely used key-value store, and permits usage of a shared-nothing, independent set of Redis nodes as a shared-something cluster, to enable intelligent partitioning of data. This is particularly helpful in use-cases that require a widely geo-distributed set of key-value store nodes, and there are main memory constraints on the hardware specifications for the nodes the key-value stores are deployed on. The experimental performance is indicative of what the system has to offer. All of the above is offered while still maintaining strong serializability guarantees for the write operations on the distributed cluster.\\

\section{Future Work}
This section lists the threads of future work that are envisioned for Redynis.

\begin{itemize}
	\item The current architecture doesn't respond well to a failure of the master propagator, on which the write serialization depends upon. Future work would primarily be focused on implementing failure handling mechanisms. These mechanisms can be introduced into the existing framework using a heartbeat mechanism to detect when the master propagator goes offline, and electing another node to take it's place.
	\item The data placement strategy that is currently in use is fairly trivial. A more sophisticated placement computation model can be plugged into RedynisDaemon, in it's stead, for more accurate placement decision strategies.
	\item The RESTful web-service wrapper in the existing implementation is only responsible to aggregating metrics and directing client requests. It can be extended to form a framework for additional data-collection, which can, in turn be used to build predictive models that can identify patterns in data accesses, and pre-emptively move data based on the features of the model learnt.\\
\end{itemize}

\section{Acknowledgments}
The author would like to thank Dr. Khuzaima Daudjee for his guidance, suggestions and feedback during the conceptualization of this research project.\\

\bibliographystyle{abbrv}
\bibliography{cs848-course-project_vineet.bib}

\begin{thebibliography}{10}

\bibitem{latency-stats}
Ip latency statistics.
\newblock \url{http://www.verizonenterprise.com/about/network/latency}.

\bibitem{redis-website}
Redis.
\newblock \url{http://redis.io}.

\bibitem{redis-popularity}
Stackoverflow developer survey.
\newblock \url{http://stackoverflow.com/research/developer-survey-2016}.

\bibitem{asad2016adaptcache}
O.~Asad and B.~Kemme.
\newblock Adaptcache: Adaptive data partitioning and migration for distributed
  object caches.
\newblock In {\em Proceedings of the 17th International Middleware Conference},
  page~7. ACM, 2016.

\bibitem{curino2010schism}
C.~Curino, E.~Jones, Y.~Zhang, and S.~Madden.
\newblock Schism: a workload-driven approach to database replication and
  partitioning.
\newblock {\em Proceedings of the VLDB Endowment}, 3(1-2):48--57, 2010.

\bibitem{redynis-daemon}
V.~John.
\newblock Redynis daemon.
\newblock \url{https://github.com/Redynis/RedynisDaemon}.

\bibitem{redynis-svc}
V.~John.
\newblock Redynis service.
\newblock \url{https://github.com/Redynis/RedynisService}.

\bibitem{pan2003overview}
J.~Pan, Y.~T. Hou, and B.~Li.
\newblock An overview of dns-based server selections in content distribution
  networks.
\newblock {\em Computer Networks}, 43(6):695--711, 2003.

\bibitem{quamar2013sword}
A.~Quamar, K.~A. Kumar, and A.~Deshpande.
\newblock Sword: scalable workload-aware data placement for transactional
  workloads.
\newblock In {\em Proceedings of the 16th International Conference on Extending
  Database Technology}, pages 430--441. ACM, 2013.

\bibitem{taft2014store}
R.~Taft, E.~Mansour, M.~Serafini, J.~Duggan, A.~J. Elmore, A.~Aboulnaga,
  A.~Pavlo, and M.~Stonebraker.
\newblock E-store: Fine-grained elastic partitioning for distributed
  transaction processing systems.
\newblock {\em Proceedings of the VLDB Endowment}, 8(3):245--256, 2014.

\end{thebibliography}

\end{document}